\documentclass[12pt]{article}

\usepackage{graphicx}

\usepackage{natbib}
\usepackage{epsfig}
\usepackage{amsfonts} 
\usepackage{amsmath}
\usepackage{amssymb,longtable,enumitem}
\usepackage{mathrsfs}
\usepackage{setspace}

\newcommand{\real}{\text{\rm I\hspace{-0.6mm}R}}  
\date{}
\LTcapwidth=\textwidth

\newcommand{\bA}{\mathbf{A}}
\newcommand{\bD}{\mathbf{D}}
\newcommand{\bI}{\mathbf{I}}
\newcommand{\bS}{\mathbf{S}}

\newcommand{\bY}{\mathbf{Y}}

\newcommand{\bmm}{\mathbf{m}}

\newcommand{\by}{\mathbf{y}}
\newcommand{\bz}{\mathbf{z}}

\newcommand{\bpi}{\boldsymbol{\pi}}

\newcommand{\btheta}{\boldsymbol{\theta}}

\newcommand{\bmu}{\boldsymbol{\mu}}
\newcommand{\bSigma}{\boldsymbol{\Sigma}}

\newcommand{\tr}{\,\mbox{tr}}
\newcommand{\diag}{\,\mbox{diag}}

\DeclareMathOperator{\Ev}{\mathbb{E}}
\DeclareMathOperator{\Var}{\mathbb{V}}

\begin{document}
\doublespacing

\title{A parsimonious family of multivariate Poisson-lognormal distributions for clustering multivariate count data}

\author{Sanjeena Subedi\footnote{Department of Mathematical Sciences, Binghamton University, State University of New York, 4400 Vestal Parkway East, Binghamton, NY, USA 13902. e: sdang@binghamton.edu} \and Ryan Browne \footnote{Department of Statistics and Actuarial Science, University of Waterloo, 200 University Avenue West, Waterloo, ON, Canada. e: ryan.browne@uwaterloo.ca}}

\maketitle

\begin{abstract}
Multivariate count data are commonly encountered through high-throughput sequencing technologies in bioinformatics, text mining, or in sports analytics. Although the Poisson distribution seems a natural fit to these count data, its multivariate extension is computationally expensive.In most cases mutual independence among the variables is assumed, however this fails to take into account the correlation among the variables usually observed in the data. Recently, mixtures of multivariate Poisson-lognormal (MPLN) models have been used to analyze such multivariate count measurements with a dependence structure. In the MPLN model, each count is modeled using an independent Poisson distribution conditional on a latent multivariate Gaussian  variable. Due to this hierarchical structure, the MPLN model can account for over-dispersion as opposed to the traditional Poisson distribution and allows for correlation between the variables. Rather than relying on a Monte Carlo-based estimation framework which is computationally inefficient, a fast variational-EM based framework is used here for parameter estimation. Further, a parsimonious family of mixtures of Poisson-lognormal distributions are proposed by decomposing the covariance matrix and imposing constraints on these decompositions. Utility of such models is shown using simulated and benchmark datasets.
\end{abstract}


\section{Introduction}
With the emergence of next generation sequencing technologies that provide a fast and cost-effective data generation platform, multivariate count data are becoming ubiquitous in bioinformatics. Although there has been a big explosion in the approaches for data generation and a dramatic reduction in cost and time, efficiently analyzing these complex biological data sets still remains a challenge. Cluster analysis is widely used in bioinformatics for identification and analysis of population heterogeneity. Clustering allows us to summarize data into homogenous groups or clusters, each with its unique characteristics. \cite{de2008} did a comparative analysis of several clustering techniques on 35 different cancer gene expression data sets and concluded that a mixture model-based clustering approach showed superior performance in terms of recovering the true structure of the data as opposed to the traditional distance-based approaches such as $k$-means and hierarchical methods. Model-based clustering, which utilizes mixture models, has been increasingly used in the last two decades  \citep{gollini2014,subedi2014, bouveyron2012, subedi2015,browne2015,kosmidis2016,dang2015,vrbik14,tortora19,subedi2020}.
A finite mixture model assumes that the population consists of a finite mixture of subpopulations or components, which can be represented by a parametric model. If  $d$-dimensional random vector $\bY$ arises from a $G$ component finite mixture model, its density can be written as 
\begin{align*}
f(\by\mid\boldsymbol{\vartheta})= \sum_{g=1}^G\pi_g f(\by\mid \boldsymbol{\varphi}_g),
\end{align*}
where $\pi_g$ for $g=1,\ldots,G$ is the mixing proportion such that $\pi_g > 0$ and $\sum_g \pi_g = 1$, $f(\cdot)$ is the component specific probability density function or the probability mass function with parameters $\boldsymbol{\varphi}_g$, and $\boldsymbol{\vartheta}$ contains all the unknown parameters for the finite mixture.  Depending on the nature of the data, an appropriate $f(\cdot)$ is used.

A number of initial works on univariate analysis for a count variable from next generation sequencing data focused on a Poisson model \citep{marioni2008,bullard2010,witten2010}. However, a Poisson distribution also imposes a mean-variance relationship such that mean and variance must be equal and most modern biological data typically exhibits over-dispersion (i.e., the observed variation is larger than what is predicted by the Poisson model, see \cite{anders2010}). A commonly used approach to incorporate this additional variability in univariate case is to model the mean parameter of Poisson distribution as a random variable and impose a distribution. Several different distributions has been proposed: the inverse-Gaussian distribution yields a Poisson-inverse Gaussian distribution \citep{holla1967}, the inverse-Gamma distribution yields a Poisson-inverse Gamma distribution \cite{willmot1993}, the lognormal distribution yields a Poisson-lognormal distribution \citep{bulmer1974}, and the gamma distribution yields a gamma-Poisson distribution (i.e., the negative binomial distribution) \citep{greenwood1920,collings1985}. See \cite{karlis2005} for a detailed list of other possible priors. Another approach is to scale the mean parameter by a random variable and impose a beta prior on that scaling random variable resulting in a Poisson-Beta distribution \citep{gurland1958}. Amongst these, the most popular approach for statistical methods pertaining next generation sequence data is the gamma-Poisson distribution also known as the negative binomial distribution \citep{robinson2010,yu2013,love2014,dong2016}. In a gamma-Poisson distribution, $\lambda$ is modeled as a random variable from the Gamma distribution such that
$Y_i\mid \lambda_i \sim \text{Poisson}(\lambda)\quad  \text{and} \quad \lambda_i \sim \text{Gamma} (\alpha,\beta)$.
The resulting marginal distribution of $Y_i$ will then be a negative binomial distribution \citep{greenwood1920,collings1985} with mean $\Ev(Y)= \alpha\beta$, dispersion parameter $\phi=\frac{1}{\alpha}$, and variance $\Var(Y)=\Ev(Y)+\Ev(Y)^2\phi >\Ev(Y)$.  Note that this is a univariate distribution. 

While the multivariate extensions of these distributions seem a natural choice to modelling multivariate count data, these models are computationally expensive and/or they impose restrictions on the covariance structure. For example, in addition to the computational burden \citep{brijs2004}, the multivariate Poisson distribution by \citep{teicher1954,campbell1934} can only allow for positive correlation among the variables \cite{karlis2007,inouye2017}. Several different formulation of multivariate version of the Poisson-beta distribution was studied by \cite{sarabia2011}; these formulations are either computationally intensive especially for high dimensional data, have restrictions on the covariance structure or the parameters in the marginal distributions are not free (i.e., they share the hyperparameters). Hence, independence between variables is assumed for multivariate analysis in most cases \citep{rau2015,si2013,papastamoulis2016}, which fails to take into account the correlation between variables. 

\cite{silva2019} developed a mixtures of multivariate Poisson-lognormal \citep[MPLN;][]{aitchison1989} distributions to cluster multivariate count measurements. In MPLN models, the counts are modeled using a hierarchical structure such that $$Y_{ij}\mid {\theta_{ij}} \sim \text{Poisson} (e^{\theta_{ij}} ) ~\text{and}~ \btheta_i =(\theta_{i1},\ldots,\theta_{id})=(\log\lambda_{i1},\ldots,\log \lambda_{id})  \sim \mathscr{N}_d (\bmu,\bSigma),$$
where $\mathscr{N}_d (\bmu,\bSigma)$ denotes a $d$-dimensional Gaussian distribution with mean $\bmu$ and covariance $\bSigma$.
Due to this hierarchical structure, the MPLN model can account for over-dispersion as opposed to the traditional Poisson distribution and allows for correlation. An attractive property of MPLN distribution is that it can support both positive and negative correlation. However, the posterior of the latent variables do not have a closed form and therefore, these models came with heavy computation cost as they relied on Bayesian MCMC-based approaches, discussed further in Section~\ref{sec:method1}. Bayesian MCMC-based approaches for models with latent variables can be computationally intensive \citep{blei2017} and the computational time is further compounded in model-based clustering where the expected value of the latent variable needs to be computed several times (further detailed in Section \ref{sec:method2}). In this paper, we propose a computationally efficient framework for clustering using mixtures of multivariate Poisson-lognormal distributions using variational Gaussian approximations. We also impose constraints on  the covariance matrices of latent variable using an eigen-decomposition which results in a parsimonious family of models. Section \ref{sec:method3} provides the variational EM estimation algorithm, Section \ref{sec:results} provides results on simulations with comparisons to a popular Poisson-based mixture approach and results on a real data set. Section \label{sec:conc} concludes with some conclusions and future directions.

 \section{Methodology}
\subsection{Multivariate Poisson-lognormal distribution}\label{sec:method1}
In multivariate Poisson-lognormal \citep[MPLN;][]{aitchison1989} distribution, the counts are modeled using a hierarchical structure such that 
\begin{align*}
Y_{ij}\mid {\theta_{ij}} \sim \text{Poisson} (e^{\theta_{ij}} ) \quad\text{and}\quad \btheta_i \sim \mathscr{N}_d (\bmu,\bSigma),
\end{align*}
where $\btheta_i= (\theta_{i1},\ldots,\theta_{id})$ and  $\mathscr{N}_d (\bmu,\bSigma)$ denotes a $d$-dimensional Gaussian distribution with mean $\bmu$ and covariance $\bSigma$. The mean and expected value of the $\bY$ \citep{aitchison1989,georgescu2011} can be found using the law of  iterative expectation,
\begin{eqnarray*}
\Ev(Y_j)=&\Ev\left[\Ev(Y_j\mid\theta_j)\right]&=e^{\mu_j+1/2 \times \Sigma_{jj}},\\
  \Var(Y_j)=& \Ev\left[\Var(Y_j\mid\theta_j)\right]+\Var\left[\Ev(Y_j\mid\theta_j)\right]&= \Ev(Y_j)+ \Ev(Y_j)^2(e^{\Sigma_{jj}}-1),\\
  \text{cov}(Y_j,Y_{k})=&\Ev\left[\text{cov}(Y_j,Y_k\mid\theta_j,\theta_k)\right]+\text{cov}\left[\Ev(Y_j\mid\theta_j)\Ev(Y_k\mid\theta_k)\right]&= \Ev(Y_j) \Ev(Y_k)(e^{\Sigma_{jk}}-1),
\end{eqnarray*}
where $Y_j$ is the $j^{th}$ entry of $\bY$, $\mu_j$ is the $j^{th}$ entry of $\bmu$, $\Sigma_{jk}$ is the entry in the $j^{th}$ row and $k^{th}$ column of the matrix $\bSigma$ and $j\neq k$. Due to this hierarchical structure, the MPLN model can account for over-dispersion and also allows for correlation. 

The marginal probability density of $\bY$ can be written as:
\vspace{-0.05in}
\begin{align*}
 f_{\bY}(\by)=\int_{\real^d} \left[ \prod _{j=1}^d p(y_{j}\mid\theta_{j})\right] ~\phi_{d}(\btheta\mid\bmu,\bSigma)~d\btheta,
\end{align*}
where $\theta_j$ and $y_j$ are the $j^{th}$ element of $\btheta$ and $\by$ respectively, $p(\cdot)$ is the probability mass function of the Poisson distribution with mean $\lambda_j=e^{\theta_j}$ and $\phi_d(\cdot)$ is the probability density function of  $d$-dimensional Gaussian distribution with mean $\bmu$ and covariance $\bSigma$. Note that the marginal distribution of $Y$ involves multiple integrals and cannot be further simplified. Hence, \cite{aitchison1989} suggested that the maximum likelihood estimation of the parameters requires a mix of Newton-Raphson and steepest ascent methods. An alternate approach is to use an expectation-maximization (EM) algorithm \citep{dempster77}, which is an iterative approach for maximizing the likelihood when the data are incomplete or are treated as incomplete. In an MPLN distribution, the observed variables are the counts $\bY$ and the missing data are the latent variables $\btheta$. The EM-algorithm comprises of two iterative steps: an E-step and an M-step. In E-step, the expected value of the complete data (i.e. observed and missing data) log-likelihood given the observed data and current parameter estimate (i.e., updates from previous iteration) is computed which is then maximized in the subsequent M-step. These step are repeated until convergence to obtain the maximum likelihood estimate of the parameters. In order to compute the expected value of the complete data log-likelihood, $\Ev(\btheta\mid \by)$ and $\Ev(\btheta\btheta'\mid\by)$ need to be computed. The conditional distribution of $f_{\btheta\mid \bY}(\btheta\mid\by,\bmu,\bSigma)$ is given by
\begin{align*}
f_{\btheta\mid \bY}(\btheta\mid\by,\bmu,\bSigma)=\frac{\left[ \prod _{j=1}^d p(y_{j} \mid \theta_{j})\right]~\phi_d(\btheta\mid\bmu,\bSigma)}{ f_{\bY}(\by\mid \bmu,\bSigma)}. 
\end{align*}
However, as mentioned above, the marginal distribution of $Y$ involves multiple integrals and cannot be further simplified. \cite{georgescu2011} proposed a maximum likelihood estimation of the parameters for MPLN distribution in the Monte Carlo EM (MCEM) framework \citep{wei1990}; they utilized a Monte Carlo EM algorithm using importance sampling to find the empirical expected value of the complete data log-likelihood. \cite{chagneau2011} utilized an MCMC-based approach for parameter estimation. However, it comes with a heavy computational overhead, further exacerbated as the dimensionality and sample size rise \citep{chagneau2011}.

\subsection{Mixtures of Multivariate Poisson-lognormal distributions}\label{sec:method2}
 \cite{silva2019} proposed a mixture of multivariate Poisson-lognormal distributions to cluster multivariate count measurements. A $G$-component mixture of MPLN distributions can be written as
\vspace{-0.1in} 
\begin{align*}
 f(\by\mid\boldsymbol{\vartheta})=\sum_{g=1}^G\pi_gf_{\bY}(\by\mid\bmu_g,\bSigma_g) = \sum_{g=1}^G\pi_g\int_{\real^d} \left[ \prod _{j=1}^d p(y_{ij}\mid \theta_{ij})\right] ~\phi_d(\btheta_i\mid\bmu_g,\bSigma_g)~d\btheta_i,
\end{align*}
where $\boldsymbol{\vartheta}$ denotes all model parameters and $f_{\bY}(\by\mid\bmu_g,\bSigma_g)$ denotes the distribution of the $g^{th}$ component with parameters $\bmu_g$ and $\bSigma_g$. In model-based clustering, an additional component membership indicator variable $\mathbf{Z}$ is introduced which is assumed to be unknown and $Z_{ig}=1$ if the observation $i^{th}$ belongs to group $g$ and $Z_{ig}=0$ otherwise. Hence, the complete data now comprises of observed expression levels $\by$, underlying latent variable $\btheta$, and unknown group membership $\mathbf{z}$. Therefore, the complete-data log-likelihood is
\begin{align*}
 l_c(\bmu,\bSigma \mid \by,\btheta,\bz) = \sum_{g=1}^G\sum_{i=1}^n z_{ig}\left[\log\phi_d(\btheta_i\mid\bmu_g,\bSigma_g) + \sum_{j=1}^d \log  p(y_{ij}\mid \theta_{ij})\right].  
\end{align*}

In order to utilize the EM framework for parameter estimation for the mixtures of MPLN distributions, finding the expected value of the complete-data log-likelihood requires the conditional expectations $\Ev(Z_{ig}\btheta_i\mid \by_i,\bmu_g,\bSigma_g)$ and $\Ev(Z_{ig}\btheta_i\btheta_i'\mid\by_i,\bmu_g,\bSigma_g)$. \cite{silva2019} utilized an MCMC-EM algorithm using \texttt{Stan} \citep{hoffman2014,Stan} to compute the required conditional expectations. {\sf Stan} uses the No-U-Turn Sampler \citep[NUTS;][]{hoffman2014}, an adaptive variant of the Hamiltonian Monte Carlo \citep[HMC;][]{neal2011} that requires no tuning parameters and can efficiently sample from a posterior distribution when the parameters are correlated. However, in a clustering context, even such efficient sampling techniques like NUTS comes with heavy computational burden as MCMC-EM needs to be conducted at every iteration of MCMC-EM after updating the model parameters and the component indicator variable. Additionally, Bayesian techniques based on MCMC sampling algorithms also comes with increased computational overhead and possible difficulty in determining convergence as the complexity of the model increases. When dealing with such complex models, convergence can also be quite slow. This crucial aspect is partly why applications of such models have been limited to small dimensions \citep{georgescu2011}. Additionally, when the true number of groups, i.e., the number of components of the mixture model, is unknown, the MCMC-EM algorithm must be employed for every possible number of components and the optimal number of components is chosen in conjunction with a model selection criterion which might entail extreme computational cost.

\subsection{Variational approximation of mixtures of MPLN distributions}\label{sec:method3}
\noindent Variational  approximation \citep{wainwright2008} is an approximate inference technique which has been very popular in machine learning. It presents an alternative parameter estimation framework for MPLN distribution by using a computationally convenient approximating density in place of a more complex but `true' posterior density. Using computationally convenient Gaussian densities, complex posterior distributions are approximated by minimizing the Kullback-Leibler (KL) divergence between the true and the approximating densities and therefore, reducing the computational overhead. Several studies have shown that it delivers accurate approximations \citep{archambeau2007,challis2013,arridge2018}. \cite{khan2013} developed variational inference framework for several non-conjugate latent Gaussian models. \cite{arridge2018} also developed a variational Gaussian approximation (VGA) to the posterior distribution arising from the univariate Poisson model with a Gaussian prior  and derived an explicit expression for the lower bound, and showed the existence and uniqueness of the optimal Gaussian approximation. 

Here, we propose a variational Gaussian approximation to the marginal distribution of the observed variable i.e. $f(\by)$ in the mixtures of MPLN distributions. Suppose, we have an approximating density $q(\btheta)$, the marginal log-likelihood can be written as
\begin{align*}
\log f(\by) &= \int_{\real^d} \log f(\by) q(\btheta) d\btheta\\
&=  \int_{\real^d} \log \frac{f(\by,\btheta)/q(\btheta)}{f(\btheta\mid\by)/q(\btheta)} q(\btheta) d\btheta\\
&= \int_{\real^d} \left[\log f(\by,\btheta) - \log q(\btheta)\right] q(\btheta)  d\btheta + D_{KL}(q\|f) \\
&= F(q,\by) + D_{KL}(q\|f),
\end{align*}

\noindent where $D_{KL}(q\|f)= \int_{\real^d} q(\btheta) \log \frac{q(\btheta)}{f(\btheta\mid\by)} d\btheta$ is the Kullback-Leibler (KL) divergence between $f(\btheta\mid \by)$ and approximating distribution $q(\btheta)$, and $F(q,\by)=\int_{\real^d} \left[\log f(\by,\btheta) - \log q(\btheta)\right] q(\btheta)  d\btheta$ is our evidence lower bound (ELBO). To minimize the KL divergence, we maximize our ELBO.
In VGA, $q(\btheta)$ is assumed to be a Gaussian distribution. Assuming $q(\btheta) = \mathscr{N}(\mathbf{m}, \mathbf{S})$, the lower bound for the log $f(\by)$ becomes
\begin{align*}
F(q,\by) &=-\frac{1}{2}(\bmm-\bmu)'\bSigma^{-1}(\bmm-\bmu)-\frac{1}{2} \tr(\bSigma^{-1}\mathbf{S})+\frac{1}{2} \log |\mathbf{S}| -\frac{1}{2} \log |\bSigma| - \frac{d}{2}+ \bmm'\by \\
&-   \sum_{j=1}^d \left(e^{m_{j}+\frac{1}{2}S_{jj}}+\log(y_{j}!)\right).
\end{align*}

This lower bound is strictly jointly concave with respect to the mean ($\bmm$) and variance ($\mathbf{S}$) of the approximating distribution and hence, similar to \cite{arridge2018}, parameter estimation can be obtained via Newton's method and fixed-point method. This alleviates the need of the use of MCMC-based approach to get samples from the posterior distribution of the latent variable $\btheta$ and the computational burden that comes along with such approaches. Additionally, the EM framework using variational Gaussian approximation algorithm is monotonic as opposed to MCMC-based EM. A similar variational framework for MPLN distribution was recently proposed by \cite{Chiquet2019} for network models. \par

The complete data log-likelihood of the mixtures of MPLN distributions can be written as:
\begin{align*}
l_c(\boldsymbol{\vartheta}\mid \by) &=\sum_{g=1}^G \sum_{i=1}^n z_{ig} \log \pi_g + \sum_{g=1}^G \sum_{i=1}^n z_{ig} \log f(\by_i\mid \bmu_g,\bSigma_g) \\
&= \sum_{g=1}^G \sum_{i=1}^n z_{ig} \log \pi_g +  \sum_{g=1}^G \sum_{i=1}^n z_{ig} \left[  \int_{\real^d} \left[\log f(\by,\btheta_{ig}) - \log q(\btheta_{ig})\right] q(\btheta_{ig})  d\btheta_{ig} + D_{KL}(q_{ig}\|f_{ig}) \right]
\end{align*}
 \noindent where $D_{KL}(q_{ig}\|f_{ig})= \int_{\real^d} q(\btheta_{ig}) \log \frac{q(\btheta_{ig})}{f(\btheta_{i}\mid\by_i,Z_{ig}=1)} d\btheta_{ig}$ is the Kullback-Leibler (KL) divergence between $f(\btheta_i\mid \by_i,Z_{ig}=1)$ and approximating distribution $q(\btheta_{ig})$. Assuming $q(\btheta_{ig})=\mathscr{N}(\bmm_{ig},\bS_{ig})$, the log-likelihood becomes:
 \begin{align*}
l_c(\boldsymbol{\vartheta}\mid \by) 
&= \sum_{g=1}^G \sum_{i=1}^n z_{ig} \log \pi_g +  \sum_{g=1}^G \sum_{i=1}^n z_{ig} \left[  F(q_{ig},\by_i)  + D_{KL}(q_{ig}\|f_{ig}) \right],
\end{align*}
 where the ELBO for each observation $\by_i$ is
 \begin{align*}
F(q_{ig},\by_i) &=\frac{1}{2} \log |\mathbf{S}_{ig}|-\frac{1}{2}(\bmm_{ig}-\bmu_g)'\bSigma_g^{-1}(\bmm_{ig}-\bmu_g)-\tr(\bSigma_g^{-1}\mathbf{S}_{ig}) -\frac{1}{2} \log |\bSigma_g| - \frac{d}{2} + \bmm_{ig}'\by_i  \\
&-  \sum_{j=1}^d \left(e^{m_{igj}+\frac{1}{2}S_{ig,jj}}+\log(y_{ij}!)\right),
\end{align*}
where $m_{igj}$ is the $j^{th}$ element of the $\bmm_{ig}$ and $S_{ig,jj}$ is the $j^{th}$ diagonal element of the matrix $\mathbf{S}_{ig}$.

The variational parameters that maximize the ELBO will minimize the KL divergence between the true posterior and the approximating density.

Parameter estimation can be done in an iterative EM-type approach such that the following steps are iterated.
\begin{enumerate}[noitemsep]
\item Conditional on the variational parameters $\bmm_{ig},\mathbf{S}_{ig}$ and on $\bmu_g$ and $\bSigma_g$, the  $\Ev(Z_{ig})$ is computed. Given $\bmu_g$ and $\bSigma_g$, 
\begin{align*}
\Ev(Z_{ig}\mid \by_i)=\frac{\pi_g f(\by\mid \bmu_g,\bSigma_g)}{\sum_{h=1}^G\pi_h f(\by\mid \bmu_h,\bSigma_h)}.
\end{align*}
Note that this involves the marginal distribution of $Y$ which is difficult to compute. Hence, we use an approximation of $\Ev(Z_{ig})$ where we replace the marginal density of the exponent of ELBO such that 
\begin{align*}
\widehat{Z}_{ig}\stackrel{\text{def}}{=}\frac{\pi_g \exp\left[F\left(q_{ig},\by_i\right) \right]}{\sum_{h=1}^G\pi_h \exp\left[F\left(q_{ih},\by_i\right) \right]}.
\end{align*}
This approximation is computationally convenient and a similar framework was utilized by \cite{tang2015,gollini2014}. This approximation works well in simulation studies and real data analysis.

\item Given $\hat{Z}_{ig}$, variational parameters $\bmm_{ig}$ and $\mathbf{S}_{ig}$ is updated conditional on $\bmu_g$ and $\bSigma_g$.
The lower bound is strictly jointly concave with respect to $\bmm_{ig}$ and $\bS_{ig}$ of the approximating distribution. Therefore, we can update $\bmm_{ig}$ and $\bS_{ig}$ as following:
\begin{enumerate}
 \item fixed-point method for updating $\mathbf{S}_{ig}$ is
\begin{align*}
\hspace{-0.65in} \mathbf{S}_{ig}^{(t+1)}=\left\{\bSigma_g^{-1}+ \bI \odot \mathbf{exp}\left[\bmm_{ig}^{(t)}+\frac{1}{2} \diag \left(\mathbf{S}_{ig}^{(t)}\right)\right] \mathbf{1}_d'  \right\}^{-1}
\end{align*}
where the vector function $\mathbf{exp}\left[ \mathbf{a} \right] = (e^{a_1}, \ldots, e^{a_d})'$ is a vector of exponential each element of the $d$-dimensional vector $\mathbf{a}$, $\mbox{diag}( \mathbf{S} ) = ( \mathbf{S}_{11} \ldots,  \mathbf{S}_{dd})$ puts the diagonal elements of the $d\times d$ matrix  $\mathbf{S}$ into a d-dimensional vector, $\odot$ the Hadmard product and $\mathbf{1}_d$ is a d-dimensional vector of ones ;
\item Newton's method to update $\bmm_{ig}$ is
\begin{align*}
 \bmm_{ig}^{(t+1)}&=\bmm_{ig}^{(t)}\! -\! \mathbf{S}_{ig}^{(t+1)}\! \left\{  \mathbf{exp}\left[\bmm_{ig}^{(t)}\! +\! \frac{1}{2}\diag\left(\mathbf{S}_{ig}^{(t+1)}\right)\right]\! +\! \bSigma_g^{-1} \left(\bmm_{ig}^{(t)}-\bmu_g\right)\! -\! \by_i \right\}\!;  \quad\mbox{and}
\end{align*}
\end{enumerate}
\item given $\hat{z}_{ig}$ and the variational parameters $\bmm_{ig}$ and $\mathbf{S}_{ig}$, the parameters $\bpi$, $\bmu_g$ and $\bSigma_g$. 
\begin{align*}
\pi_g&=\frac{\sum_{i=1}^n \widehat{Z}_{ig}}{n},\\
 \bmu_g& = \frac{\sum_{i=1}^n \widehat{Z}_{ig} \bmm_{ig}}{\sum_{i=1}^n \widehat{Z}_{ig}},\quad  \text{and} \\
 \bSigma_g&=\frac{\sum_{i=1}^n \widehat{Z}_{ig} (\bmm_{ig}-\bmu_g)(\bmm_{ig}-\bmu_g)'}{\sum_{i=1}^n \widehat{Z}_{ig}}- \frac{\sum_{i=1}^n \widehat{Z}_{ig} \mathbf{S}_{ig}}{\sum_{i=1}^n \widehat{Z}_{ig}}.
\end{align*}

\end{enumerate}

\subsection{Parsimonious family of models}
For the mixtures of MPLN distribution, the number of free parameters in the covariance matrices of the latent variable is $Gd(d+1)/2$, i.e., the number of parameters increases quadratically with $d$. \cite{banfield93} proposed an eigen-decomposition of the covariance matrices of mixtures of Gaussian distributions such that $\bSigma_g = \lambda_g \bD_g\bA_g\bD_g'$ where $\bD_g$ is the matrix of eigenvectors and $\bA_g$ is a diagonal matrix proportional to the eigenvalues of $\bSigma_g$, such that $|\bA_g|=1$, and $\lambda_g$ is the associated constant of proportionality. For a Gaussian distribution, this decomposition results in a geometric interpretation of the constraints such that $\lambda_g$ controls the cluster volume, $\bA_g$ controls the cluster shape, and $\bD_g$ controls the cluster orientation. \cite{celeux95} imposed constraints on these parameters of the covariance matrix to be equal or different among groups resulting in a family of 14 models known as known as Gaussian parsimonious clustering models (GPCM). Parameter estimation in an EM framework for 12 of these 14 GPCM family and MM framework of \cite{browne14a} for the remaining two models is available in the {\sf R} package ${\tt mclust}$ \citep{scrucca2016}. Here, we propose an eigen-decomposition of the component-specific covariance matrices of the latent variable $\btheta$ to introduce parsimony. The geometric interpretation does not hold for observed data and therefore, we only focused on the following subset of the GPCM family of model given in Table \ref{eigen}. 

\begin{center}
\begin{table}[ht]
\centering
\caption{Geometric interpretation of the subset of the models considered in our study obtained via eigen-decomposition of $\bSigma_g$.\label{eigen}}
\begin{tabular*}{0.95\textwidth}{@{\extracolsep{\fill}}llllr}
\hline 
Model (Covariance) & Volume & Shape & Orientation &  Parameters\\
\hline
EII ($\lambda I$)&Equal & Spherical & -&$1$ \\
VII ($\lambda_g I$)&Variable &Spherical &-&$G$\\
EEI ($\lambda \bA$)&Equal &Equal &Ax-Alg&$d$ \\
VVI ($\lambda_g \bA_g$)& Variable&Variable&Ax-Alg&$dG$ \\
EEE ($\lambda \bD\bA\bD^{T}$)& Equal &Equal &Equal &$d(d+1)/2$\\
VVE ($\lambda_g \bD \bA_g \bD^{T}$)&Variable & Variable & Equal &$d(d+1)/2+(G-1)d$\\
EEV ($\lambda \bD_g\bA \bD_g^T$) &Equal &Equal &Variable&$Gd(d+1)/2-(G-1)d$ \\
VVV ($\lambda_g \bD_g\bA_g \bD_g^{T}$)& Variable&Variable&Variable&$Gd(d+1)/2$\\
\hline
\end{tabular*}
\end{table}
\end{center}

Note that within a component $\text{cov}(Y_j,Y_k)=\Ev(Y_j)\Ev(Y_k)(e^{\Sigma_{jk}}-1),$ therefore, independence among the latent variable $\btheta$ will result in independence among the observed variable $\bY$. We focused on EII, VII, EEI, and VVI to recover diagonal covariance structure when variables are independent. Additionally, to introduce some parsimony compared to VVV model, we also focused some additional non-diagonal constriants such as EEE, VVE and EEV. Given $\bmm_{ig}$ and $\bS_{ig}$ and using the sample covariance matrix of $\bmm_g$ as $\hat{\bSigma}_{\text{sample}}=\frac{\sum_{i=1}^n \widehat{Z}_{ig} (\bmm_{ig}-\bmu_g)(\bmm_{ig}-\bmu_g)'}{\sum_{i=1}^n \widehat{Z}_{ig}}- \frac{\sum_{i=1}^n \widehat{Z}_{ig} \mathbf{S}_{ig}}{\sum_{i=1}^n \widehat{Z}_{ig}}$ the parameter estimates are analogous to their Gaussian counterpart. See \cite{celeux95,browne14a} for details.

\subsection{Initialization, Model Selection, and Performance assessment}
EM algorithm relies heavily on the initial values \citep{biernacki2003} and hence, good starting values are crucial. While initialization of the clusters using $k$-means is one of the widely used approach for symmetric distributions, it may not be appropriate for our over-dispersed count datasets as $k$-means initialization is equivalent to fitting a Gaussian mixture models with spherical clusters which will not be ideal for these over-dispersed multivariate counts. Therefore, we utilized small EM initialization for our analysis. For each $G$, the initialization ```EII" model of consist of using 20 different random partitions of datasets into $G$ clusters and running 20 iterations of our variational EM algorithm. The parameters associated with the largest log-likelihood among the 20 random partitions was then used to initialize all eight models which is then run until convergence. Convergence of the algorithm for these models is determined using a modified Aitken acceleration criterion. 
where at iteration $m$. Convergence is based on Aitken acceleration \citep{aitken26} which is considered to satisfied when 
$l_\infty^{(m+1)}-l_\infty^{(m)} $ is positive and smaller than some $\epsilon$,
where $l^{(m)}$ is the value of the log-likelihood and  $l_\infty^{(m+1)}$ is an asymptotic estimate of the log-likelihood given by 
\begin{align*}
 l_\infty^{(m+1)} = l^{(m)}+\frac{l^{(m+1)}-l^{(m-1)}}{1-a^{(m)}}
 \qquad \mbox{where}\qquad a^{(m)}=\frac{l^{(m+1)}-l^{(m)}}{l^{(m)}-l^{(m-1)}},
\end{align*}
 \citep{bohning94}. Here, we used $\epsilon=0.001.$

Typically, the number of component is unknown in cluster analysis. In our case, the covariance structure that provides the best fit is also unknown. Hence, we fit  the model with all eight covariance structure to a range of $G$ and the Bayesian information criteria (BIC) \citep{schwarz1978} is used to assess the goodness of fit of each fitted model and select the number of $G$ and corresponding to the corresponding covariance structure that has the best fitness to the data. The BIC of a fitted model is computed by
\begin{align*}
		\mathrm{BIC(\hat{\boldsymbol{\vartheta}})} = -2\ell(\hat{\boldsymbol{\vartheta}}) + \psi \log(n),
\end{align*}
where $\ell(\boldsymbol{\vartheta})$ is the observed data log-likelihood function here, $n$ is the sample size, and $\psi$ is the number of free parameters in the model. Note that here we compute approximation of BIC using $ELBO$ \citep{chen2018}. We will refer to the approximation of BIC as BIC in the paper. Additionally, the variational parameters can be regarded as proxies for the latent variables and therefore, we do not penalize for variational parameters. The model that has the smallest BIC value is selected and the number of underlying groups is therefore estimated. Performance assessment clustering performance was done in using the Adjusted Rand Index \citep[ARI;][]{Hubert85}: 1 indicates perfect agreement between the true and predicted classification and the expected value of the ARI under random classification is zero. A value $<$ 0 indicate a classification that is worse than would be expected under random
assignment.
For simulation studies where we know the true class labels, we also compared the performance of our proposed model to a Poisson mixture model implemented in the {\sf R} package \texttt{HTSCluster} \citep{rau2015}. Note that it is a clustering algorithm designed for clustering high-throughput transcriptome sequencing (HTS) data and therefore has a built-in normalization option which could not be turned off. Therefore, we set the normalization option to ``TMM" for our analysis.

\section{Analyses}\label{sec:results}
\subsection{Simulation Study 1}
We generated 100 three dimensional datasets of size $N=2000$ with three components with mixing proportions $\boldsymbol{\pi}=(0.2,0.5,0.3)$, known mean and a completely unconstrained VVV covariance structure (see Table \ref{tabsim1} for the values used to generate the data). We ran our model for $G=1:4$ and all eight covariance structures. In 100 out of the 100 datasets, a three component VVV model was selected with an average ARI of 0.99 and standard deviation (sd) of 0.003. Summary of the parameter estimates is provided in Table \ref{tabsim1}.  
\begin{table}[!ht]
\caption{Summary of the average and standard errors (SE) of the estimated parameters from the 100 datasets where a $G=3$ component VVV model was selected in Simulation Study~1.\label{tabsim1}}
\begin{center}
	\begin{tabular*}{0.95\textwidth}{@{\extracolsep{\fill}}llll}
	\hline
	& True Parameters & Average of Estimated Parameters& Standard errors \\
	\hline
	$\bmu_1$&(6, 3, 3)&(6.00, 3.00, 3.00)&(0.03, 0.04, 0.04)\\
	$\bmu_2$&(3, 5, 3)&(3.00, 5.00, 3.00)&(0.02, 0.02, 0.02)\\
	$\bmu_3$&(5, 3, 5)&(5.00, 3.00, 5.00) &(0.02, 0.03, 0.02)\\
	\hline
\end{tabular*}
\end{center}

\begin{equation*}
\bSigma_1 =
\begin{bmatrix}
0.30 & 0.15 & 0.20 \\ 
0.15 & 0.40 & 0.30 \\ 
0.20 & 0.30 & 0.40 \\ 
\end{bmatrix}, \quad
\hat{\bSigma}_1 =
\begin{bmatrix}
0.30 & 0.15 & 0.20 \\ 
0.15 & 0.40 & 0.30 \\ 
0.20 & 0.30 & 0.40 \\ 
\end{bmatrix},
\quad
\text{SE}(\hat{\bSigma}_1)=
\begin{bmatrix}
0.02 & 0.02 & 0.02 \\ 
0.02 & 0.03 & 0.03 \\ 
0.02 & 0.03 & 0.03 \\ 
\end{bmatrix}
\end{equation*}

\begin{equation*}
\bSigma_2 =
\begin{bmatrix}
0.30 & 0.15 & 0.20 \\ 
0.15 & 0.40 & 0.30 \\ 
0.20 & 0.30 & 0.40 \\ 
\end{bmatrix}, \quad
\hat{\bSigma}_2 =
\begin{bmatrix}
 0.30 & 0.15 & 0.20 \\ 
0.15 & 0.40 & 0.30 \\ 
0.20 & 0.30 & 0.40 \\ 
\end{bmatrix},
\quad
\text{SE}(\hat{\bSigma}_2)=
\begin{bmatrix}
0.02 & 0.01 & 0.01 \\ 
0.01 & 0.02 & 0.02 \\ 
0.01 & 0.02 & 0.02 \\
\end{bmatrix}
\end{equation*}

\begin{equation*}
\bSigma_3 =
\begin{bmatrix}
0.20 & -0.15 & -0.10 \\ 
-0.15 & 0.40 & -0.10 \\ 
-0.10 & -0.10 & 0.20 \\ 
\end{bmatrix}, \quad
\hat{\bSigma}_3 =
\begin{bmatrix}
0.20 & -0.15 & -0.10 \\ 
-0.15 & 0.39 & -0.10 \\ 
-0.10 & -0.10 & 0.20 \\ 
\end{bmatrix},
\quad
\text{SE}(\hat{\bSigma}_3)=
\begin{bmatrix}
 0.01 & 0.01 & 0.01 \\ 
 0.01 & 0.02 & 0.01 \\ 
0.01 & 0.01 & 0.01 \\
\end{bmatrix}
\end{equation*}

\end{table}

Similarly, we also ran the Poisson mixture model in  the {\sf R} package \texttt{HTSCluster} for $G=1:4$ and selected the best model with BIC. In 100 out of the 100 datasets, it overestimated the number of components resulting in a four component model was selected with an average ARI of 0.75 and a standard deviation of 0.11.

\subsection{Simulation Study 2}
Here, we generated 100 six dimensional datasets of size $N=500$ with two components with mixing proportions $\boldsymbol{\pi}=(0.59,0.41)$, known mean and a completely constrained EII ($\lambda$I) covariance structure (see Table \ref{tabsim2} for the values used to generate the data). We ran our model for $G=1:3$ and all eight covariance structures. In 99 out of 100 datasets, a two component EII model was selected with an average ARI of 1.00 and a standard deviation of 0.00 and in one of the 100 dataset, a two component VII model was selected. Note, that VII is still imposes an isotropic (spherical) structure on $\bSigma$ but it allows $\bSigma$ to be vary among groups. In our case, with $G=2$, it just added one additional parameter so the penalty for selecting a slightly complicated model was very small. Summary of the parameter estimates of the 99 out of the 100 datasets where the correct model is selected is provided in \ref{tabsim2}. 

\begin{table}[!ht]
\caption{Summary of the average and standard errors (SE) of the estimated parameters from the 99 out the 100 datasets where a $G=2$ component EII model was selected in Simulation Study~2.\label{tabsim2}}
\begin{center}
\begin{tabular*}{0.9\textwidth}{@{\extracolsep{\fill}}cccc}
\hline 
g& $\bmu_g$ &  \multicolumn{2}{c}{$\hat{\bmu}_g$}  \\
	\hline
	1&(5, 6, 5, 5, 5, 6)&Average&(5.00, 6.00, 5.01, 5.00, 5.01, 6.01)\\
	&&SE&(0.05, 0.06, 0.05, 0.05, 0.05, 0.06)\\
	2&(2.5, 3, 2.5, 3, 3, 2.5)&Average&(2.49, 3.01, 2.50, 3.00, 3.00, 2.49)\\
	&&SE&(0.07, 0.07, 0.07, 0.07, 0.07, 0.07 )\\
	\hline
	g& $\lambda$ &  \multicolumn{2}{c}{$\hat{\lambda}$} \\
	\hline
	1, 2& 1& Average& 0.99\\
	&&SE&0.03\\
	\hline
\end{tabular*}
\end{center}
\end{table}

Similarly, we also ran the Poisson mixture model in  the {\sf R} package \texttt{HTSCluster} for $G=1:3$ and selected the best model with BIC. In 100 out of the 100 datasets, it overestimated the number of components resulting in a three component model was selected with an average ARI of 0.04 and a standard deviation of 0.02.

\subsection{Simulation studies on data generated from other models}
\subsubsection{Simulation Study 3: Datasets generated using mixtures of negative binomial distributions}
Here, we generated 100 six dimensional datasets of size $N=2000$ from a two component mixtures of independent negative binomial distributions with mixing proportions $\boldsymbol{\pi}=(0.79,0.21)$. We ran all our eight models for $G=1,\ldots,4$ and model selection was done using BIC. In 100 out of 100 datasets, a two EII component model was selected with perfect classification. Note that, it always selected a diagonal covariance structure where the covariance is 0.
Table \ref{tabsim3neg} shows that our model was able to recover the true mean and variance of the data generated from the mixtures of negative binomial distributions.

\begin{table}[!ht]
\caption{Summary of the average and standard errors (SE) of the estimated means for the 100 datasets in Simulation Study~3 where the data was generated from a mixture of independent negative binomial distributions.\label{tabsim3neg}}
\begin{center}
\begin{tabular*}{0.99\textwidth}{@{\extracolsep{\fill}}ll}
\hline
 \multicolumn{2}{c}{\textbf{g=1}}\\
\hline
True Mean &(1000, 500, 1000, 500, 1000 , 500)\\
Average of Estimated Means & (1000.28, 500.24, 999.84, 500.09, 999.75, 500.25)\\
SE of Estimated Means&(2.66, 1.50, 2.31, 1.43, 2.91, 1.36)\\
\hline
True Variance&(11000, 3000, 11000, 3000, 11000, 3000)\\
Average of Estimated Variances & (11045.99, 3012.67, 11036.90, 30011.06, 11034.85, 3012.86)\\
SE of Estimated Variances&(147.40, 37.25, 148.18, 37.31, 152.14, 39.56)\\
\hline
 \multicolumn{2}{c}{\textbf{g=2}}\\
\hline
True Mean &(500, 1000, 500, 1000, 500, 500)\\
Average of Estimated Means & (500.10, 999.81,  500.00, 1000.53, 499.91, 499.75)\\
SE of Estimated Means&(2.44, 5.38, 5.37, 2.71, 2.39, 2.68)\\
\hline 
True Variance&(3000, 11000, 3000,11000, 3000, 3000)\\
Average of Estimated Variances & (3011.23, 11036.40, 3010.06, 11051.85, 3009.10, 3007.34)\\
SE of Estimated Variances&(43.19, 182.23, 45.03, 193.22, 43.65, 43.70)\\
\hline
\end{tabular*}
\end{center}
\end{table}

Similarly, we also ran the Poisson mixture model in  the {\sf R} package \texttt{HTSCluster} for $G=1:4$ and selected the best model with BIC. In 100 out of the 100 datasets, it overestimated the number of components resulting in a four component model was selected with an average ARI of 0.29 and a standard deviation of 0.01.

\subsubsection{Simulation Study 4: Datasets generated using mixtures of independent Poisson distributions}
Here, we used 100 four dimensional datasets generated using a mixtures of independent Poisson distributions with mixing proportions $\bpi=(0.59,0.41)$.We ran all our eight models for $G=1,\ldots,4$ and model selection was done using BIC. In 100 out of the 100 datasets, a two component EII model was selected with perfect classification for all datasets. Note that here, the diagonal structure for $\bSigma$ was always selected as the datasets were generated from mixtures of independent Poisson distributions. Table \ref{tabsim3pois} shows that our model was able to recover the true mean the data generated from the mixtures of Poisson distributions very well but it overestimated the variable slightly which would be expected as MPLN imposes a structure such that $\Var(Y)>\Ev(Y)$ but for Poisson distribution $\Var(Y)=\Ev(Y).$ 

\begin{table}[!ht]
\caption{Summary of the average and standard errors (SE) of the estimated means for the 100 datasets in Simulation Study~4 where the data was generated using a mixtures of independent Poisson distributions.\label{tabsim3pois}}
\begin{center}
\begin{tabular*}{0.9\textwidth}{@{\extracolsep{\fill}}ll}
\hline
 \multicolumn{2}{c}{\textbf{g=1}}\\
\hline
True Mean &(1000, 1500, 1500, 1000)\\
Average of Estimated Means &(999.98, 1500.16, 1500.27, 1000.09)\\
SE of Estimated Means&(1.31, 1.84, 1.77, 1.40)\\
\hline
True Variance&(1000, 1500, 1500, 1000)\\
Average of Estimated Variances &(1027.97, 1563.15, 1563.27,1028.09)\\
SE of Estimated Variances&(6.21, 14.37, 14.28, 6.52)\\
\hline
 \multicolumn{2}{c}{\textbf{g=2}}\\
\hline
True Mean &(1000, 1000, 1000, 1500)\\
Average of Estimated Means &(999.94, 1000.04, 999.96, 1499.97) \\
SE of Estimated Means&(1.39, 1,32, 1.38, 1.79)\\
\hline 
True Variance&(1000, 1000, 1000, 1500)\\
Average of Estimated Variances &(1027.93, 1028.03, 1027.95, 1562.95 ) \\
SE of Estimated Variances&(6.50, 6.13, 6.57, 14.51)\\
\hline
\end{tabular*}
\end{center}
\end{table}
Similarly, we also ran the Poisson mixture model in  the {\sf R} package \texttt{HTSCluster} for $G=1:4$ and selected the best model with BIC. In 100 out of the 100 datasets, it selected the correct $G=2$ model with an ARI of 1.

\subsection{Real data analysis}
\subsection{Chronic Kidney Disease Data Set}
Here, we used the Chronic Kidney Disease (CKD) Dataset available via UCI Machine Learning repository. The dataset comprises of 25 attributes regarding various blood measurements and various categorical measurements pertaining disease status on 400 individuals. Here, we focus on three discrete count measurements: \texttt{bgr}: blood glucose random; \texttt{wbcc}: white blood cell count; and \texttt{pcv}packed cell volume with the aim of classifying the disease status of the patient for CKD. Information on the status of CKD is available in the variable \texttt{class}. Before the analysis, observations with missing values on our 4 variables (\texttt{bgr, wbcc, pcv} and \texttt{class}) were removed resulting in 261 observations. We ran our algorithm for $G=1,\ldots,4$ and BIC selected a two component "VVI" model. As seen in Table \ref{ckdtab}, the first cluster comprised of mostly CKD patients and the second cluster comprised mostly of non-CKD patients. 

\begin{table}[!ht]
\caption{Cross tabulation of the clusters obtained by mixtures of MPLN and HTSCluster against the CKD status.\label{ckdtab}}
\begin{center}
\begin{tabular*}{0.9\textwidth}{@{\extracolsep{\fill}}llllllll}
\hline
& \multicolumn{2}{c}{Mix. MPLN}&&\multicolumn{4}{c}{HTSCluster}\\\cline{2-3}\cline{5-8}
&1&2&&1&2&3&4\\
\hline
CKD&95 & 29 & &28&18&29&49\\
Not CKD & 0 & 135& & 24&0&16&95\\
\hline
\end{tabular*}
\end{center}
\end{table}

\begin{figure}[ht]
\begin{center}
\includegraphics[width=0.32\textwidth]{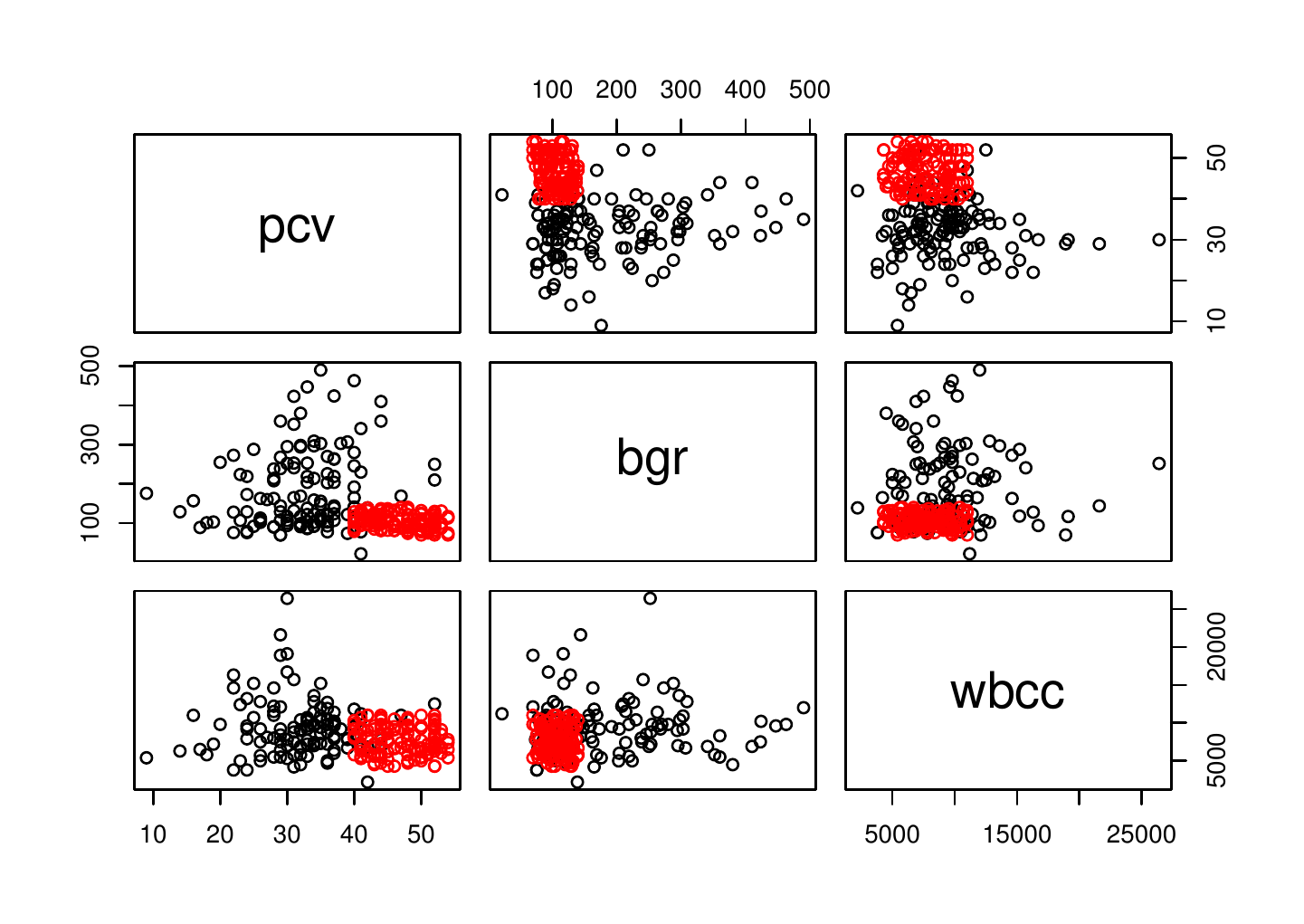} 
\includegraphics[width=0.32\textwidth]{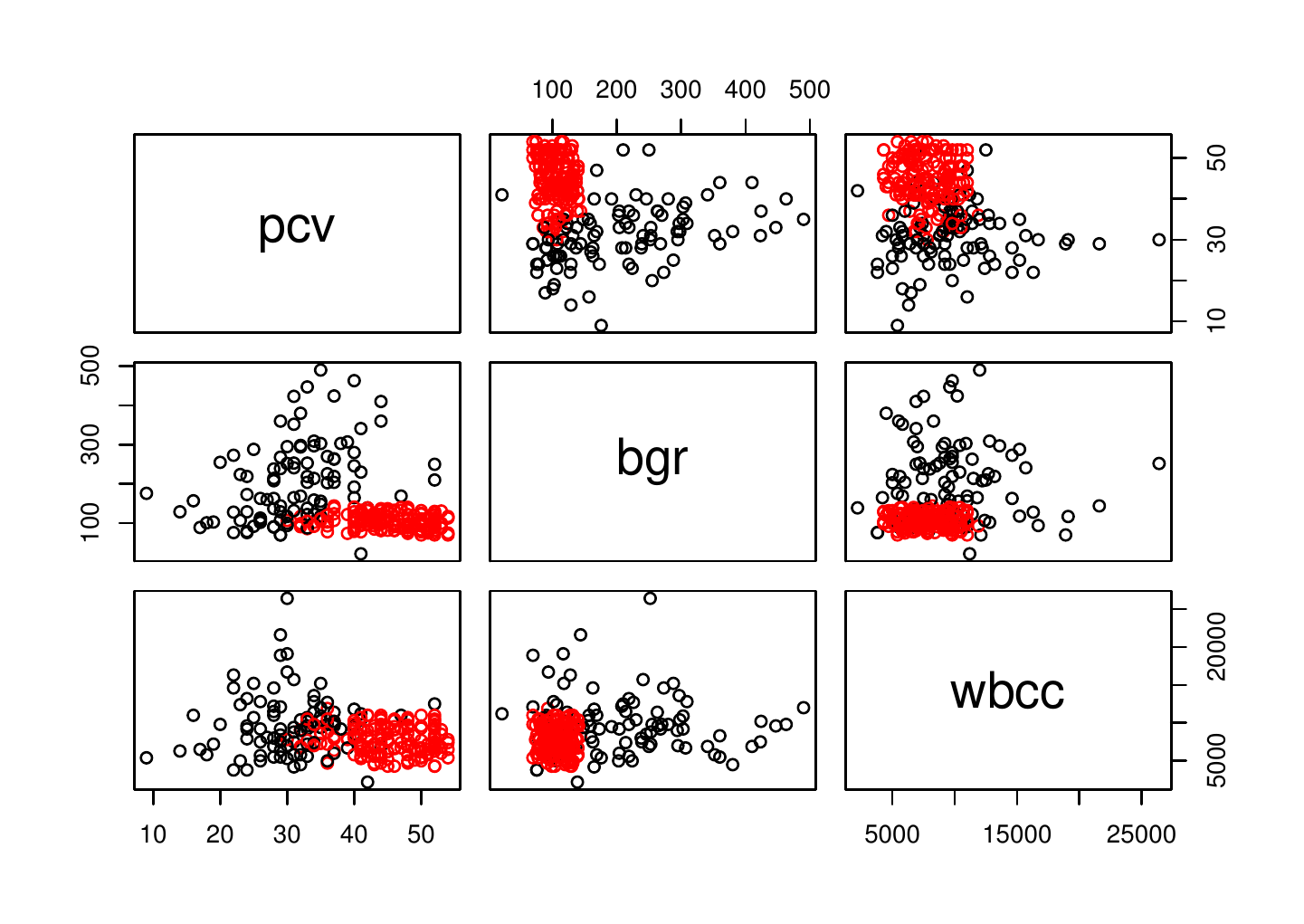} 
\includegraphics[width=0.32\textwidth]{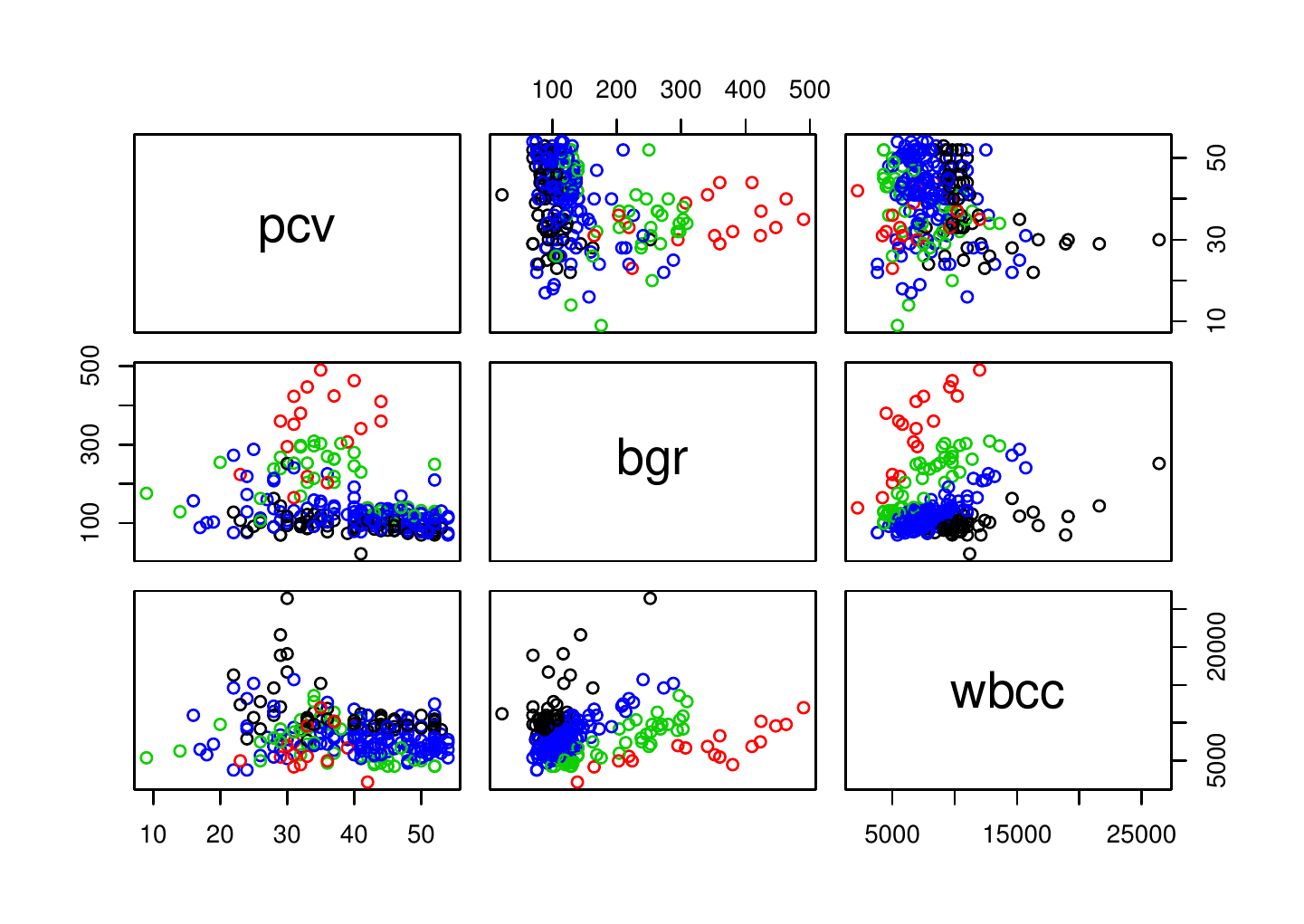} 
\caption{Scatter plot matrix of the chronic kidney dataset where the colors/symbols represents the CKD status on the left, colors/symbols representing the cluster structure recovered by MPLN in the middle, and colors/symbols representing the cluster structure recovered by HTSCluster on the right.}
\label{CKD}
\end{center}
\end{figure}

Visualization of the true cluster structure and predicted cluster structure are shown in Figure \ref{CKD}. We also ran HTSCluster on the dataset for $G=1,\ldots,4$ and it selected a four component model. However, the cluster structure appears to be random (see Figure \ref{CKD} and Table \ref{ckdtab}).

\section{Conclusions}\label{sec:conc}
Here, we propose an efficient framework for parameter estimation for clustering using mixtures of multivariate Poisson-lognormal distributions that utilizes a variational Gaussian approximation. The variational Gaussian approximation provides a fast and deterministic framework for estimating the model parameters and alleviates the computational overhead due to MCMC-EM. Through eigen-decomposition of the component's covariance matrices of the latent variable and imposition of constraints, a family of parsimonious models is proposed. Through simulation studies, we show that the proposed models provide competitive performance, the parameters were very close to the true parameters (when the correct model was chosen), and close to perfect classification was obtained using the model selected by BIC. Additionally, we also generate data from mixtures of univariate Poisson distributions and mixtures of negative binomial distributions and show that our proposed model can recover the true mean of the models fairly well. When datasets were generated from a Poisson model which assumes  $\Var(Y)=\Ev(Y)$, the variance was overestimated slightly as MPLN imposes the structure that $\Var(Y)>\Ev(Y).$ In the case where the datasets were generated from a mixture of univariate negative binomial distributions, the variance estimation was much closer as both MPLN and negative binomial impose the structure that $\Var(Y)>\Ev(Y).$  Additionally, when data are generated from mixtures of univariate distributions, our approach selects a model with a diagonal covariance structure where variables are independent.

While our proposed work is motivated by modern biological data such as RNA-seq data, here we only focus on a general framework that is applicable to any multivariate count data. For biological data, extensions of these approaches may require some structure that needs to be imposed on the datasets to normalize for library specific factors such as library size and transcript specific factors such as gene length. Although the use of variational Gaussian approximations alleviates some of the challenges of MCMC-EM, the algorithm can still be prone to computational issues encountered by any traditional EM algorithm such as convergence to local maxima, singularities, etc. Furthermore, a systematic evaluation for the speed and accuracy of variational EM against the MCMC-EM would be very valuable and is ongoing. In our work, BIC seems to perform fairly well, however, some future work will focus on investigation of different model-selection criteria for MPLN mixture models and for discrete data in general.


\begin{thebibliography}{}

\bibitem[\protect\citeauthoryear{Aitchison and Ho}{Aitchison and
  Ho}{1989}]{aitchison1989}
Aitchison, J. and C.~Ho (1989).
\newblock The multivariate {P}oisson-log normal distribution.
\newblock {\em Biometrika\/}~{\em 76\/}(4), 643--653.

\bibitem[\protect\citeauthoryear{Aitken}{Aitken}{1926}]{aitken26}
Aitken, A.~C. (1926).
\newblock A series formula for the roots of algebraic and transcendental
  equations.
\newblock {\em Proceedings of the Royal Society of Edinburgh\/}~{\em 45},
  14--22.

\bibitem[\protect\citeauthoryear{Anders and Huber}{Anders and
  Huber}{2010}]{anders2010}
Anders, S. and W.~Huber (2010).
\newblock Differential expression analysis for sequence count data.
\newblock {\em Nature Precedings\/}, 1--1.

\bibitem[\protect\citeauthoryear{Archambeau, Cornford, Opper, and
  Shawe-Taylor}{Archambeau et~al.}{2007}]{archambeau2007}
Archambeau, C., D.~Cornford, M.~Opper, and J.~Shawe-Taylor (2007).
\newblock Gaussian process approximations of stochastic differential equations.
\newblock {\em Journal of Machine Learning Research\/}~{\em 1}, 1--16.

\bibitem[\protect\citeauthoryear{Arridge, Ito, Jin, and Zhang}{Arridge
  et~al.}{2018}]{arridge2018}
Arridge, S.~R., K.~Ito, B.~Jin, and C.~Zhang (2018).
\newblock Variational gaussian approximation for poisson data.
\newblock {\em Inverse Problems\/}~{\em 34\/}(2), 025005.

\bibitem[\protect\citeauthoryear{Banfield and Raftery}{Banfield and
  Raftery}{1993}]{banfield93}
Banfield, J.~D. and A.~E. Raftery (1993).
\newblock Model-based {G}aussian and non-{G}aussian clustering.
\newblock {\em Biometrics\/}~{\em 49\/}(3), 803--821.

\bibitem[\protect\citeauthoryear{Biernacki, Celeux, and Govaert}{Biernacki
  et~al.}{2003}]{biernacki2003}
Biernacki, C., G.~Celeux, and G.~Govaert (2003).
\newblock Choosing starting values for the em algorithm for getting the highest
  likelihood in multivariate gaussian mixture models.
\newblock {\em Computational Statistics \& Data Analysis\/}~{\em 41\/}(3-4),
  561--575.

\bibitem[\protect\citeauthoryear{Blei, Kucukelbir, and McAuliffe}{Blei
  et~al.}{2017}]{blei2017}
Blei, D.~M., A.~Kucukelbir, and J.~D. McAuliffe (2017).
\newblock Variational inference: A review for statisticians.
\newblock {\em Journal of the American statistical Association\/}~{\em
  112\/}(518), 859--877.

\bibitem[\protect\citeauthoryear{B\"{o}hning, Dietz, Schaub, Schlattmann, and
  Lindsay}{B\"{o}hning et~al.}{1994}]{bohning94}
B\"{o}hning, D., E.~Dietz, R.~Schaub, P.~Schlattmann, and B.~Lindsay (1994).
\newblock The distribution of the likelihood ratio for mixtures of densities
  from the one-parameter exponential family.
\newblock {\em Annals of the Institute of Statistical Mathematics\/}~{\em 46},
  373--388.

\bibitem[\protect\citeauthoryear{Bouveyron and Brunet}{Bouveyron and
  Brunet}{2012}]{bouveyron2012}
Bouveyron, C. and C.~Brunet (2012).
\newblock Simultaneous model-based clustering and visualization in the fisher
  discriminative subspace.
\newblock {\em Statistics and Computing\/}~{\em 22\/}(1), 301--324.

\bibitem[\protect\citeauthoryear{Brijs, Karlis, Swinnen, Vanhoof, Wets, and
  Manchanda}{Brijs et~al.}{2004}]{brijs2004}
Brijs, T., D.~Karlis, G.~Swinnen, K.~Vanhoof, G.~Wets, and P.~Manchanda (2004).
\newblock A multivariate poisson mixture model for marketing applications.
\newblock {\em Statistica Neerlandica\/}~{\em 58\/}(3), 322--348.

\bibitem[\protect\citeauthoryear{Browne and McNicholas}{Browne and
  McNicholas}{2014}]{browne14a}
Browne, R.~P. and P.~D. McNicholas (2014).
\newblock Estimating common principal components in high dimensions.
\newblock {\em Advances in Data Analysis and Classification\/}~{\em 8\/}(2),
  217--226.

\bibitem[\protect\citeauthoryear{Browne and McNicholas}{Browne and
  McNicholas}{2015}]{browne2015}
Browne, R.~P. and P.~D. McNicholas (2015).
\newblock A mixture of generalized hyperbolic distributions.
\newblock {\em Canadian Journal of Statistics\/}~{\em 43\/}(2), 176--198.

\bibitem[\protect\citeauthoryear{Bullard, Purdom, Hansen, and Dudoit}{Bullard
  et~al.}{2010}]{bullard2010}
Bullard, J.~H., E.~Purdom, K.~D. Hansen, and S.~Dudoit (2010).
\newblock Evaluation of statistical methods for normalization and differential
  expression in mrna-seq experiments.
\newblock {\em BMC Bioinformatics\/}~{\em 11\/}(1), 94.

\bibitem[\protect\citeauthoryear{Bulmer}{Bulmer}{1974}]{bulmer1974}
Bulmer, M. (1974).
\newblock On fitting the poisson lognormal distribution to species-abundance
  data.
\newblock {\em Biometrics\/}, 101--110.

\bibitem[\protect\citeauthoryear{Campbell}{Campbell}{1934}]{campbell1934}
Campbell, J. (1934).
\newblock The poisson correlation function.
\newblock {\em Proceedings of the Edinburgh Mathematical Society\/}~{\em
  4\/}(1), 18--26.

\bibitem[\protect\citeauthoryear{Celeux and Govaert}{Celeux and
  Govaert}{1995}]{celeux95}
Celeux, G. and G.~Govaert (1995).
\newblock Gaussian parsimonious clustering models.
\newblock {\em Pattern Recognition\/}~{\em 28}, 781--793.

\bibitem[\protect\citeauthoryear{Chagneau, Mortier, Picard, and Bacro}{Chagneau
  et~al.}{2011}]{chagneau2011}
Chagneau, P., F.~Mortier, N.~Picard, and J.-N. Bacro (2011).
\newblock A hierarchical bayesian model for spatial prediction of multivariate
  non-gaussian random fields.
\newblock {\em Biometrics\/}~{\em 67\/}(1), 97--105.

\bibitem[\protect\citeauthoryear{Challis and Barber}{Challis and
  Barber}{2013}]{challis2013}
Challis, E. and D.~Barber (2013).
\newblock Gaussian kullback-leibler approximate inference.
\newblock {\em The Journal of Machine Learning Research\/}~{\em 14\/}(1),
  2239--2286.

\bibitem[\protect\citeauthoryear{Chen, Wang, Erosheva, et~al.}{Chen
  et~al.}{2018}]{chen2018}
Chen, Y.-C., Y.~S. Wang, E.~A. Erosheva, et~al. (2018).
\newblock On the use of bootstrap with variational inference: Theory,
  interpretation, and a two-sample test example.
\newblock {\em The Annals of Applied Statistics\/}~{\em 12\/}(2), 846--876.

\bibitem[\protect\citeauthoryear{Chiquet, Mariadassou, and Robin}{Chiquet
  et~al.}{2019}]{Chiquet2019}
Chiquet, J., M.~Mariadassou, and S.~Robin (2019).
\newblock Variational inference for sparse network reconstruction from count
  data.
\newblock In {\em Proceedings of the 36th International Conference on Machine
  Learning}, Volume~97 of {\em Proceedings of Machine Learning Research}.

\bibitem[\protect\citeauthoryear{Collings and Margolin}{Collings and
  Margolin}{1985}]{collings1985}
Collings, B.~J. and B.~H. Margolin (1985).
\newblock Testing goodness of fit for the poisson assumption when observations
  are not identically distributed.
\newblock {\em Journal of the American Statistical Association\/}~{\em
  80\/}(390), 411--418.

\bibitem[\protect\citeauthoryear{Dang, Browne, and McNicholas}{Dang
  et~al.}{2015}]{dang2015}
Dang, U.~J., R.~P. Browne, and P.~D. McNicholas (2015).
\newblock Mixtures of multivariate power exponential distributions.
\newblock {\em Biometrics\/}~{\em 71\/}(4), 1081--1089.

\bibitem[\protect\citeauthoryear{de~Souto, Costa, de~Araujo, Ludermir, and
  Schliep}{de~Souto et~al.}{2008}]{de2008}
de~Souto, M.~C., I.~G. Costa, D.~S. de~Araujo, T.~B. Ludermir, and A.~Schliep
  (2008).
\newblock Clustering cancer gene expression data: a comparative study.
\newblock {\em BMC Bioinformatics\/}~{\em 9\/}(1), 1.

\bibitem[\protect\citeauthoryear{Dempster, Laird, and Rubin}{Dempster
  et~al.}{1977}]{dempster77}
Dempster, A.~P., N.~M. Laird, and D.~B. Rubin (1977).
\newblock Maximum likelihood from incomplete data via the {EM} algorithm.
\newblock {\em Journal of the Royal Statistical Society: Series B\/}~{\em
  39\/}(1), 1--38.

\bibitem[\protect\citeauthoryear{Dong, Zhao, Tong, and Wan}{Dong
  et~al.}{2016}]{dong2016}
Dong, K., H.~Zhao, T.~Tong, and X.~Wan (2016).
\newblock Nblda: negative binomial linear discriminant analysis for rna-seq
  data.
\newblock {\em BMC Bioinformatics\/}~{\em 17\/}(1), 369.

\bibitem[\protect\citeauthoryear{Georgescu, Desassis, Soubeyrand, Kretzschmar,
  and Senoussi}{Georgescu et~al.}{2014}]{georgescu2011}
Georgescu, V., N.~Desassis, S.~Soubeyrand, A.~Kretzschmar, and R.~Senoussi
  (2014).
\newblock An automated {MCEM} algorithm for hierarchical models with
  multivariate and multitype response variables.
\newblock {\em Communications in Statistics-Theory and Methods\/}~{\em
  43\/}(17), 3698--3719.

\bibitem[\protect\citeauthoryear{Gollini and Murphy}{Gollini and
  Murphy}{2014}]{gollini2014}
Gollini, I. and T.~B. Murphy (2014).
\newblock Mixture of latent trait analyzers for model-based clustering of
  categorical data.
\newblock {\em Statistics and Computing\/}~{\em 24\/}(4), 569--588.

\bibitem[\protect\citeauthoryear{Greenwood and Yule}{Greenwood and
  Yule}{1920}]{greenwood1920}
Greenwood, M. and G.~U. Yule (1920).
\newblock An inquiry into the nature of frequency distributions representative
  of multiple happenings with particular reference to the occurrence of
  multiple attacks of disease or of repeated accidents.
\newblock {\em Journal of the Royal statistical society\/}~{\em 83\/}(2),
  255--279.

\bibitem[\protect\citeauthoryear{Gurland}{Gurland}{1958}]{gurland1958}
Gurland, J. (1958).
\newblock A generalized class of contagious distributions.
\newblock {\em Biometrics\/}~{\em 14\/}(2), 229--249.

\bibitem[\protect\citeauthoryear{Hoffman and Gelman}{Hoffman and
  Gelman}{2014}]{hoffman2014}
Hoffman, M.~D. and A.~Gelman (2014).
\newblock The no-u-turn sampler: adaptively setting path lengths in hamiltonian
  monte carlo.
\newblock {\em Journal of Machine Learning Research\/}~{\em 15\/}(1),
  1593--1623.

\bibitem[\protect\citeauthoryear{Holla}{Holla}{1967}]{holla1967}
Holla, M. (1967).
\newblock On a poisson-inverse gaussian distribution.
\newblock {\em Metrika\/}~{\em 11\/}(1), 115--121.

\bibitem[\protect\citeauthoryear{Hubert and Arabie}{Hubert and
  Arabie}{1985}]{Hubert85}
Hubert, L. and P.~Arabie (1985).
\newblock Comparing partitions.
\newblock {\em Journal of Classification\/}~{\em 2}, 193--218.

\bibitem[\protect\citeauthoryear{Inouye, Yang, Allen, and Ravikumar}{Inouye
  et~al.}{2017}]{inouye2017}
Inouye, D.~I., E.~Yang, G.~I. Allen, and P.~Ravikumar (2017).
\newblock A review of multivariate distributions for count data derived from
  the poisson distribution.
\newblock {\em Wiley Interdisciplinary Reviews: Computational
  Statistics\/}~{\em 9\/}(3), e1398.

\bibitem[\protect\citeauthoryear{Karlis and Meligkotsidou}{Karlis and
  Meligkotsidou}{2007}]{karlis2007}
Karlis, D. and L.~Meligkotsidou (2007).
\newblock Finite mixtures of multivariate poisson distributions with
  application.
\newblock {\em Journal of Statistical Planning and Inference\/}~{\em 137\/}(6),
  1942--1960.

\bibitem[\protect\citeauthoryear{Karlis and Xekalaki}{Karlis and
  Xekalaki}{2005}]{karlis2005}
Karlis, D. and E.~Xekalaki (2005).
\newblock Mixed poisson distributions.
\newblock {\em International Statistical Review\/}~{\em 73\/}(1), 35--58.

\bibitem[\protect\citeauthoryear{Khan, Aravkin, Friedlander, and Seeger}{Khan
  et~al.}{2013}]{khan2013}
Khan, M.~E., A.~Aravkin, M.~Friedlander, and M.~Seeger (2013).
\newblock Fast dual variational inference for non-conjugate latent gaussian
  models.
\newblock In {\em International Conference on Machine Learning}, pp.\
  951--959.

\bibitem[\protect\citeauthoryear{Kosmidis and Karlis}{Kosmidis and
  Karlis}{2016}]{kosmidis2016}
Kosmidis, I. and D.~Karlis (2016).
\newblock Model-based clustering using copulas with applications.
\newblock {\em Statistics and Computing\/}~{\em 26\/}(5), 1079--1099.

\bibitem[\protect\citeauthoryear{Love, Huber, and Anders}{Love
  et~al.}{2014}]{love2014}
Love, M.~I., W.~Huber, and S.~Anders (2014).
\newblock Moderated estimation of fold change and dispersion for rna-seq data
  with deseq2.
\newblock {\em Genome Biology\/}~{\em 15\/}(12), 550.

\bibitem[\protect\citeauthoryear{Marioni, Mason, Mane, Stephens, and
  Gilad}{Marioni et~al.}{2008}]{marioni2008}
Marioni, J.~C., C.~E. Mason, S.~M. Mane, M.~Stephens, and Y.~Gilad (2008).
\newblock Rna-seq: an assessment of technical reproducibility and comparison
  with gene expression arrays.
\newblock {\em Genome Research\/}~{\em 18\/}(9), 1509--1517.

\bibitem[\protect\citeauthoryear{Neal et~al.}{Neal et~al.}{2011}]{neal2011}
Neal, R.~M. et~al. (2011).
\newblock Mcmc using hamiltonian dynamics.
\newblock {\em Handbook of Markov Chain Monte Carlo\/}~{\em 2\/}(11).

\bibitem[\protect\citeauthoryear{Papastamoulis, Martin-Magniette, and
  Maugis-Rabusseau}{Papastamoulis et~al.}{2016}]{papastamoulis2016}
Papastamoulis, P., M.-L. Martin-Magniette, and C.~Maugis-Rabusseau (2016).
\newblock On the estimation of mixtures of {P}oisson regression models with
  large number of components.
\newblock {\em Computational Statistics \& Data Analysis\/}~{\em 93}, 97--106.

\bibitem[\protect\citeauthoryear{Rau, Maugis-Rabusseau, Martin-Magniette, and
  Celeux}{Rau et~al.}{2015}]{rau2015}
Rau, A., C.~Maugis-Rabusseau, M.-L. Martin-Magniette, and G.~Celeux (2015).
\newblock Co-expression analysis of high-throughput transcriptome sequencing
  data with {P}oisson mixture models.
\newblock {\em Bioinformatics\/}~{\em 31\/}(9), 1420--1427.

\bibitem[\protect\citeauthoryear{Robinson, McCarthy, and Smyth}{Robinson
  et~al.}{2010}]{robinson2010}
Robinson, M.~D., D.~J. McCarthy, and G.~K. Smyth (2010).
\newblock edger: a bioconductor package for differential expression analysis of
  digital gene expression data.
\newblock {\em Bioinformatics\/}~{\em 26\/}(1), 139--140.

\bibitem[\protect\citeauthoryear{Sarabia and G{\'o}mez-D{\'e}niz}{Sarabia and
  G{\'o}mez-D{\'e}niz}{2011}]{sarabia2011}
Sarabia, J.~M. and E.~G{\'o}mez-D{\'e}niz (2011).
\newblock Multivariate poisson-beta distributions with applications.
\newblock {\em Communications in Statistics-Theory and Methods\/}~{\em
  40\/}(6), 1093--1108.

\bibitem[\protect\citeauthoryear{Schwarz et~al.}{Schwarz
  et~al.}{1978}]{schwarz1978}
Schwarz, G. et~al. (1978).
\newblock Estimating the dimension of a model.
\newblock {\em The Annals of Statistics\/}~{\em 6\/}(2), 461--464.

\bibitem[\protect\citeauthoryear{Scrucca, Fop, Murphy, and Raftery}{Scrucca
  et~al.}{2016}]{scrucca2016}
Scrucca, L., M.~Fop, T.~B. Murphy, and A.~E. Raftery (2016).
\newblock {mclust} 5: clustering, classification and density estimation using
  {G}aussian finite mixture models.
\newblock {\em The {R} Journal\/}~{\em 8\/}(1), 205--233.

\bibitem[\protect\citeauthoryear{Si, Liu, Li, and Brutnell}{Si
  et~al.}{2014}]{si2013}
Si, Y., P.~Liu, P.~Li, and T.~P. Brutnell (2014).
\newblock Model-based clustering for {RNA}-seq data.
\newblock {\em Bioinformatics\/}~{\em 30\/}(2), 197--205.

\bibitem[\protect\citeauthoryear{Silva, Rothstein, McNicholas, and
  Subedi}{Silva et~al.}{2019}]{silva2019}
Silva, A., S.~J. Rothstein, P.~D. McNicholas, and S.~Subedi (2019).
\newblock A multivariate poisson-log normal mixture model for clustering
  transcriptome sequencing data.
\newblock {\em BMC Bioinformatics\/}~{\em 20\/}(1), 394.

\bibitem[\protect\citeauthoryear{{Stan Development Team}}{{Stan Development
  Team}}{2015}]{Stan}
{Stan Development Team} (2015).
\newblock {\em {Stan: A C++ library for probability and sampling (Version
  2.8.0)}}.

\bibitem[\protect\citeauthoryear{Subedi and McNicholas}{Subedi and
  McNicholas}{2014}]{subedi2014}
Subedi, S. and P.~D. McNicholas (2014).
\newblock Variational bayes approximations for clustering via mixtures of
  normal inverse gaussian distributions.
\newblock {\em Advances in Data Analysis and Classification\/}~{\em 8\/}(2),
  167--193.

\bibitem[\protect\citeauthoryear{Subedi and McNicholas}{Subedi and
  McNicholas}{2020}]{subedi2020}
Subedi, S. and P.~D. McNicholas (2020).
\newblock A variational approximations-dic rubric for parameter estimation and
  mixture model selection within a family setting.
\newblock {\em Journal of Classification\/}, 1--20.

\bibitem[\protect\citeauthoryear{Subedi, Punzo, Ingrassia, and
  McNicholas}{Subedi et~al.}{2015}]{subedi2015}
Subedi, S., A.~Punzo, S.~Ingrassia, and P.~D. McNicholas (2015).
\newblock Cluser-weighed $t$-facor analyzers for robus model-based clusering
  and dimension reducion.
\newblock {\em Statistical Methods \& Applications\/}~{\em 24\/}(4), 623--649.

\bibitem[\protect\citeauthoryear{Tang, Browne, and McNicholas}{Tang
  et~al.}{2015}]{tang2015}
Tang, Y., R.~P. Browne, and P.~D. McNicholas (2015).
\newblock Model based clustering of high-dimensional binary data.
\newblock {\em Computational Statistics \& Data Analysis\/}~{\em 87}, 84--101.

\bibitem[\protect\citeauthoryear{Teicher}{Teicher}{1954}]{teicher1954}
Teicher, H. (1954).
\newblock On the multivariate poisson distribution.
\newblock {\em Scandinavian Actuarial Journal\/}~{\em 1954\/}(1), 1--9.

\bibitem[\protect\citeauthoryear{Tortora, Franczak, Browne, and
  McNicholas}{Tortora et~al.}{2019}]{tortora19}
Tortora, C., B.~C. Franczak, R.~P. Browne, and P.~D. McNicholas (2019).
\newblock A mixture of coalesced generalized hyperbolic distributions.
\newblock {\em Journal of Classification\/}~{\em 36\/}(1), 26--57.

\bibitem[\protect\citeauthoryear{Vrbik and McNicholas}{Vrbik and
  McNicholas}{2014}]{vrbik14}
Vrbik, I. and P.~D. McNicholas (2014).
\newblock Parsimonious skew mixture models for model-based clustering and
  classification.
\newblock {\em Computational Statistics and Data Analysis\/}~{\em 71},
  196--210.

\bibitem[\protect\citeauthoryear{Wainwright, Jordan, et~al.}{Wainwright
  et~al.}{2008}]{wainwright2008}
Wainwright, M.~J., M.~I. Jordan, et~al. (2008).
\newblock Graphical models, exponential families, and variational inference.
\newblock {\em Foundations and Trends{\textregistered} in Machine
  Learning\/}~{\em 1\/}(1--2), 1--305.

\bibitem[\protect\citeauthoryear{Wei and Tanner}{Wei and
  Tanner}{1990}]{wei1990}
Wei, G.~C. and M.~A. Tanner (1990).
\newblock A {M}onte {C}arlo implementation of the {EM} algorithm and the poor
  man's data augmentation algorithms.
\newblock {\em Journal of the American Statistical Association\/}~{\em
  85\/}(411), 699--704.

\bibitem[\protect\citeauthoryear{Willmot}{Willmot}{1993}]{willmot1993}
Willmot, G.~E. (1993).
\newblock On recursive evaluation of mixed poisson probabilities and related
  quantities.
\newblock {\em Scandinavian Actuarial Journal\/}~{\em 1993\/}(2), 114--133.

\bibitem[\protect\citeauthoryear{Witten, Tibshirani, Gu, Fire, and Lui}{Witten
  et~al.}{2010}]{witten2010}
Witten, D., R.~Tibshirani, S.~G. Gu, A.~Fire, and W.-O. Lui (2010).
\newblock Ultra-high throughput sequencing-based small rna discovery and
  discrete statistical biomarker analysis in a collection of cervical tumours
  and matched controls.
\newblock {\em BMC Biology\/}~{\em 8\/}(1), 58.

\bibitem[\protect\citeauthoryear{Yu, Huber, and Vitek}{Yu
  et~al.}{2013}]{yu2013}
Yu, D., W.~Huber, and O.~Vitek (2013).
\newblock Shrinkage estimation of dispersion in negative binomial models for
  rna-seq experiments with small sample size.
\newblock {\em Bioinformatics\/}~{\em 29\/}(10), 1275--1282.

\end{thebibliography}


\end{document}